\documentclass[reprint,
superscriptaddress,
 amsmath,amssymb,
 aps,
pra,
showkeys,
floatfix,
]{revtex4-2}
\usepackage{graphicx}
\usepackage{dcolumn}
\usepackage{bm}
\usepackage{tensor}
\usepackage[dvipsnames]{xcolor}
\usepackage{soul}
\usepackage{svg}
\usepackage{subcaption}
\usepackage{tabularray}
\usepackage[stretch=10,shrink=10]{microtype}
\usepackage{physics}
\usepackage{amsmath}
\usepackage[utf8]{inputenc} 
\usepackage[T1]{fontenc}
\usepackage{dsfont}
\usepackage[symbol]{footmisc}
\usepackage[%
    colorlinks=true,
    pdfborder={0 0 0},
    citecolor=blue,
    linkcolor=blue
]{hyperref}
\usepackage{url}
\begin{document}
\setstcolor{red}
\newcommand{\XW}[1]{{\color{blue}{[XW: #1]}}}
\newcommand{\ANC}[1]{{\color{red}{[ANC: #1]}}}
\newcommand{\YJ}[1]{{\color{teal}{ #1}}}
\newcommand{\rs}[1]{\setstcolor{red}\st{#1}}       

\preprint{APS/123-QED}
\title{Mitigating cosmic ray-like correlated events with a modular quantum processor}

\author{Xuntao Wu}
\thanks{These two authors contributed equally.}
\affiliation{Pritzker School of Molecular Engineering, University of Chicago, Chicago, IL 60637, USA}

\author{Yash J. Joshi}
\thanks{These two authors contributed equally.}
\affiliation{Pritzker School of Molecular Engineering, University of Chicago, Chicago, IL 60637, USA}

\author{Haoxiong Yan}
\altaffiliation[Present address: ]{Applied Materials Inc, Santa Clara, CA 95051, USA}
\affiliation{Pritzker School of Molecular Engineering, University of Chicago, Chicago, IL 60637, USA}

\author{Gustav Andersson}
\affiliation{Pritzker School of Molecular Engineering, University of Chicago, Chicago, IL 60637, USA}

\author{Alexander Anferov}
\altaffiliation[Present address: ]{Department of Physics, ETH Zürich, 8093 Zürich, Switzerland}
\affiliation{Pritzker School of Molecular Engineering, University of Chicago, Chicago, IL 60637, USA}

\author{Christopher R. Conner}
\affiliation{Pritzker School of Molecular Engineering, University of Chicago, Chicago, IL 60637, USA}

\author{Bayan Karimi}
\affiliation{Pritzker School of Molecular Engineering, University of Chicago, Chicago, IL 60637, USA}

\author{Amber M. King}
\affiliation{Pritzker School of Molecular Engineering, University of Chicago, Chicago, IL 60637, USA}

\author{Shiheng Li}
\affiliation{Pritzker School of Molecular Engineering, University of Chicago, Chicago, IL 60637, USA}
\affiliation{Department of Physics, University of Chicago, Chicago, IL 60637, USA}

\author{Howard L. Malc}
\affiliation{Pritzker School of Molecular Engineering, University of Chicago, Chicago, IL 60637, USA}

\author{Jacob M. Miller}
\affiliation{Pritzker School of Molecular Engineering, University of Chicago, Chicago, IL 60637, USA}
\affiliation{Department of Physics, University of Chicago, Chicago, IL 60637, USA}

\author{Harsh Mishra}
\affiliation{Pritzker School of Molecular Engineering, University of Chicago, Chicago, IL 60637, USA}

\author{Hong Qiao}
\affiliation{Pritzker School of Molecular Engineering, University of Chicago, Chicago, IL 60637, USA}

\author{Minseok Ryu}
\affiliation{Pritzker School of Molecular Engineering, University of Chicago, Chicago, IL 60637, USA}

\author{Siyuan Xing}
\affiliation{Pritzker School of Molecular Engineering, University of Chicago, Chicago, IL 60637, USA}
\affiliation{Department of Mathematics, University of Chicago, Chicago, IL 60637, USA}

\author{Jian Shi}
\affiliation{Pritzker School of Molecular Engineering, University of Chicago, Chicago, IL 60637, USA}
\affiliation{Department of Materials Science and Engineering, Rensselaer Polytechnic Institute, Troy, NY 12180, USA}

\author{Andrew N. Cleland}
\email{anc@uchicago.edu}
\affiliation{Pritzker School of Molecular Engineering, University of Chicago, Chicago, IL 60637, USA}
\affiliation{Center for Molecular Engineering and Material Science Division, Argonne National Laboratory, Lemont, IL 60439, USA}

\date{\today}

\begin{abstract}
Quantum processors based on superconducting qubits are being scaled to larger qubit numbers, enabling the implementation of small-scale quantum error correction codes. However, catastrophic chip-scale correlated errors have been observed in these processors, attributed to e.g. cosmic ray impacts, which challenge conventional error-correction codes such as the surface code. These events are characterized by a temporary but pronounced suppression of the qubit energy relaxation times. Here, we explore the potential for modular quantum computing architectures to mitigate such correlated energy decay events. We measure cosmic ray-like events in a quantum processor comprising a motherboard and two flip-chip bonded daughterboard modules, each module containing two superconducting qubits. We monitor the appearance of correlated qubit decay events within a single module and across the physically separated modules. We find that while decay events within one module are strongly correlated (over $85\%$), events in separate modules only display $\sim 2\%$ correlations. We also report coincident decay events in the motherboard and in either of the two daughterboard modules, providing further insight into the nature of these decay events. These results suggest that modular architectures, combined with bespoke error correction codes, offer a promising approach for protecting future quantum processors from chip-scale correlated errors.
\end{abstract}

\maketitle
\section{Introduction}
Fault-tolerant quantum computing enables the implementation of complex quantum algorithms through the use of quantum error correction (QEC) codes \cite{Shor1996, Preskill1997, Bombin2006, Gottesman2009, Fowler2012, Bravyi2024}. However, conventional QEC codes, which have been implemented in some small-scale quantum systems~\cite{Kelly2015, Gong2021, Chen2021, Marques2021, Krinner2022, Zhao2022QEC, Acharya2023, Sivak2023, Ni2023, Bluvstein2023, Gupta2024, Cai2024, Caune2024, Acharya2024, Besedin2025, Putterman2025}, are known to be susceptible to correlated errors appearing in physical qubits~\cite{Fowler2014}, with variant codes developed for some specific types of correlated errors~\cite{Pattison2023,Tiurev2023, Kam2024}.

The recent observations of chip-scale correlated events, mostly attributed to cosmic rays or other external sources of high-energy particles~\cite{Vepslinen2020, Cardani2023, Loer2024}, present an important challenge for superconducting quantum systems \cite{Cardani2021, Wilen2021, McEwen2021, Thorbeck2023, Harrington2024, Li2024, Bratrud2024, Dedominicis2024}. These cosmic ray-like (CRL) interactions are thought to generate large numbers of THz-frequency acoustic phonons, which when interacting with a superconducting metal, generate populations of quasiparticles (QPs), poisoning the superconducting qubits \cite{Martinis2009, Catelani2011, Liu2024}. The signature for these events is a strong reduction in the qubit energy relaxation time $T_1$ \cite{Martinis2021} due to tunneling of QPs across the qubit junction. Decay events of this type are highly correlated across multiple qubits on the same substrate, resulting in the loss of quantum states stored in the qubits, which conventional QEC codes are unable to protect.

Efforts to reduce the impact of these decay events include better shielding techniques \cite{Vepslinen2020, Cardani2021, Cardani2023, Bratrud2024}, constructing devices with multiple superconducting bandgap energies \cite{Marchegiani2022, Pan2022, Kamenov2024, McEwen2024, Yelton2025, Nho2025}, substrate backside metalization for phonon absorption~\cite{Iaia2022, Larson2025}, normal-metal quasiparticle traps~\cite{Riwar2016}, and noise environment engineering~\cite{Gustavsson2016}. Although some approaches have reduced the rate of catastrophic decay events, correlated events with the attendant loss of quantum states still occur. Here, we explore the potential for a modular superconducting quantum computing architecture~\cite{Jiang2007, Zhong2021, Gold2021, Niu2023, Storz2023, Zhou2023, Field2024, Mollenhauer2024, Wu2024, Norris2025, Almanakly2025} to confine and reduce the impact of CRL decay events. 

Our modular quantum processor \cite{Wu2024} comprises two daughterboard modules, each containing two transmon qubits, flip-chip assembled on a common motherboard, the motherboard including a four-way dynamic router that provides all-to-all qubit connectivity. The qubits on the daughterboards are non-galvanically coupled to the router elements and other control and readout circuitry on the motherboard. This design happens to isolate QPs generated in the daughterboards or motherboard from the other modular elements. As the physical contact area between the daughterboards and the motherboard is small and made of soft amorphous polymers, phonon transmission between module elements is also impeded. We find that this modular design localizes decay events to each module, significantly reducing correlated events across modules. By combining standard QEC codes with higher-level codes, this localization provides a means to protect distributed quantum states from catastrophic loss in CRL events \cite{Xu2022}.

\section{Experimental setup}

\subsection{Physical platform}
\begin{figure}[tb]
    \centering
    \includegraphics[width=\linewidth]{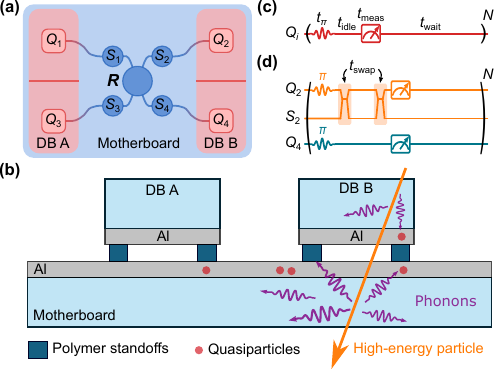}
    \caption{\label{fig:1} Layout and measurement pulse sequence. (a) Schematic for the modular quantum processor. Two daughterboards (DB), labeled A (B), each supporting two qubits $Q_{1,3}$ ($Q_{2,4}$), are flip-chip bonded with soft polymeric materials to the motherboard, which supports the router $R$ with its four switches $S_j$ as well as control and readout circuitry. Qubits on each daughterboard are galvanically separate from each other, as indicated by the solid red lines in the middle of both red shaded areas. (b) Schematic cross section of assembly (not to scale), with superconducting Al films (grey), polymer standoffs (dark blue), and sapphire substrates (light blue). A high-energy particle (orange) creates phonons (purple) in the sapphire substrates, which in turn generate quasiparticles (red) in the superconducting films. (c) Pulse sequence to detect CRL events in a qubit: We repeatedly apply a 50 ns $\pi$ pulse to the qubit, then following a delay $t_{\rm idle} = 1~\mu$s, measure the qubit during $t_{\rm meas} = 800$~ns. After a wait time $t_{\rm wait} = 8.15~\mu$s, the sequence is repeated, with a total cycle time of $10~\mu$s. (d) Pulse sequence to detect CRL events in the motherboard, by swapping an excitation from $Q_2$ into and out of the router via switch $S_2$ during $t_{\mathrm{swap}}=3.6~\mathrm{ns}$ with an intermediate wait time $t_{\mathrm{idle}}=1~\mu\mathrm{s}$, then measure $Q_2$. This is done in parallel with a CRL-detecting sequence on other qubits (here $Q_4$).}
\end{figure}

We perform our measurements on a composite superconducting device described previously \cite{Wu2024}, comprising a 20 mm $\times$ 20 mm motherboard die hosting control and readout wiring as well as an externally controlled quantum router, and two physically separate 3 mm $\times$ 7 mm daughterboard modules, flip-chip bonded to the motherboard using polymer adhesives and standoffs \cite{Satzinger2019, Conner2021}, shown schematically in Fig.~\ref{fig:1}(a) and (b). There are no galvanic connections between the daughterboard modules or between either daughterboard and the motherboard. Each daughterboard module A (B) includes two flux-tunable superconducting transmon qubits \cite{Barends2013} $Q_{1,3}$ ($Q_{2,4}$), capacitively coupled to router switches $S_{1,3}$ ($S_{2,4}$) on the motherboard, where each switch is a flux-tunable SQUID coupler \cite{Wu2024}. Circuit wiring is defined by patterning a $100$ nm thick layer of Al on a sapphire die and Josephson junctions are formed by overlapping a thick electrode ($100$ nm) on a thin oxidized electrode ($32.5$ nm) via a Dolan bridge technique~\cite{Dolan1977}. The switches are connected in common to a central capacitor, forming the externally-controlled router $R$. The router provides user-selectable connections between the qubits, allowing pairwise, three-way, and all-to-all connections between the four qubits in the two daughterboard modules. We note that the qubits on the same daughterboard are completely galvanically isolated from one another, with the superconducting ground planes severed at the midplane of each daughterboard module. The switch elements on the motherboard are in galvanic contact with each other via the central capacitor, as well as to wiring that provides capacitive coupling to each qubit, but are galvanically isolated from the motherboard ground plane. The assembled device is placed inside an aluminum enclosure, with aluminum wire-bond connections between the motherboard and a circuit board. The enclosure is mounted with the plane of the device oriented horizontally on the mixing chamber plate of a dilution refrigerator that operates below $10$ mK. We note the experiment is in the second basement of a building with five stories at and above grade, with an estimated total concrete thickness of $220$ cm above the experiment.

We operate the device as shown in Fig.~\ref{fig:1}, with an example CRL event shown schematically in Fig.~\ref{fig:1}(b) and CRL-detecting measurement sequences in Fig.~\ref{fig:1}(c) and (d). Qubit $Q_3$ was not operational; with the exception of the switch $S_2$ dynamically connected to $Q_2$, we set the fluxes in the switches $S_{1, 3, 4}$ to turn them off and isolate the qubits. High-energy particles interact with daughterboards and the motherboard, generating superconducting pair-breaking phonons which in turn create QPs in the superconducting films, localized in the daughterboards and motherboards by the non-galvanic connections. The resulting sudden increase in QP population reduces the corresponding qubit energy lifetimes $T_1$, detected in the daughterboards by the measurement sequence in Fig.~\ref{fig:1}(c) and in the motherboard using the sequence in Fig.~\ref{fig:1}(d).

\subsection{Measurement scheme\label{sec:measurement_scheme}}
We separately optimize the qubit readout visibilities using the daughterboard readout resonators. After characterizing the main qubit parameters (see Appendix~\ref{sec:qubitparams}), we simultaneously monitor qubit decay events in qubits $Q_1$, $Q_2$, and $Q_4$, using the measurement sequence in Fig.~\ref{fig:1}(c). We record each qubit measurement result as a ``$g$'' (ground state) or ``$e$'' (excited state), then analyze the recorded sequence in a manner similar to Ref.~\cite{Harrington2024}. The data reported here represent a total of $5.87$ hours of observation time, not including data communication and processing interruptions; this corresponds to $2.64 \times 10^9$ measurements of each qubit.

In a separate set of measurements, we also monitor the router on the motherboard. As we cannot directly measure the router, we use $Q_2$ as an ancilla, using the sequence in Fig.~\ref{fig:1}(d): After $Q_2$ is excited, we tune $Q_2$ and $S_2$ into resonance for a swap time $t_{\mathrm{swap}}=3.6~\mathrm{ns}$. We then tune $S_2$ to its idle point, wait for $t_{\mathrm{idle}}=1~\mu\mathrm{s}$, then retune $S_2$ and $Q_2$ into resonance and swap any remaining excitation to $Q_2$, followed by a $Q_2$ measurement. Note that our router lifetime is measured to be around $10.1~\mu\mathrm{s}$ (see Appendix~\ref{sec:router_lifetime}), sufficiently long compared to $t_{\rm idle}$ such that, in the absence of a CRL event, the router will most likely (over $90\%$) stay in its excited state. This sequence is in tandem with CRL-detection measurements of $Q_1$ and $Q_4$. In this set of measurements, we accumulated $5.7$ hours of observation data, corresponding to $2.48 \times 10^9$ measurements for each qubit.

If we measure a qubit in $g$, the excitation and measurement of that qubit in the subsequent cycle is expected to yield $e$, given the short idle time $t_{\rm idle}$ compared to the typical qubit $T_1$. Two consecutive $g$s ($gg$) indicate that an error has occurred, which could be due to state preparation and measurement (SPAM) errors or a CRL event. To reduce the effect of SPAM errors we extend this to a template $ggg$, post-processing the data to be ``$1$'' when this template sequence is observed and ``$0$'' otherwise. When a CRL event occurs, the qubit $T_1$ is typically suppressed for a few ms, resulting in a long, mostly uninterrupted series of $g$s, and a corresponding series of post-processed ``1''s, as illustrated in Fig.~\ref{fig:4}(a) and (b). We accumulate data into time bins of $N_b = 100$ measurement cycles, and convert the number $T$ of occurrences of the template sequence into a time bin decay probability $P_d = T/T_{\rm max}$, where $T_{\rm max} = N_b - 2$.

\section{Experimental results}

\begin{figure}[tb]
    \centering
    \includegraphics[width=\linewidth]{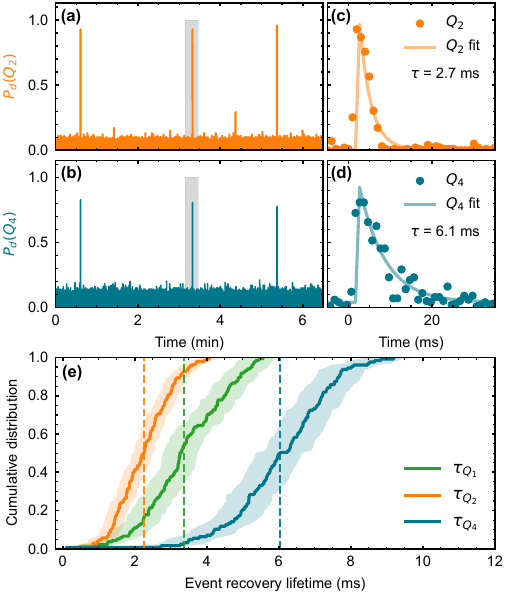}
    \caption{Short sequences of binned data for qubits $Q_2$ and $Q_4$. (a), (b) Decay probabilities for $Q_2$ and $Q_4$ averaged in $N_b = 100$ cycles. (c), (d) Detailed time dependence for coincident events at $t = 3~\mathrm{min}$ (shaded in (a) and (b)), with an exponential fit yielding the event recovery time $\tau$. (e) Cumulative distribution of event recovery times for qubits $Q_{1, 2, 4}$;  shaded regions represent the fit uncertainty of one standard deviation. Vertical dashed lines mark the mean values, with $\overline{\tau_{Q_1}} = 3.7 \pm 0.1$ ms, $\overline{\tau_{Q_2}} = 2.2 \pm 0.1$ ms, and $\overline{\tau_{Q_4}} = 6.2 \pm 0.1$ ms.} 
    \label{fig:2}
\end{figure}

\begin{figure*}[tb]
    \centering
    \includegraphics[width=\textwidth]{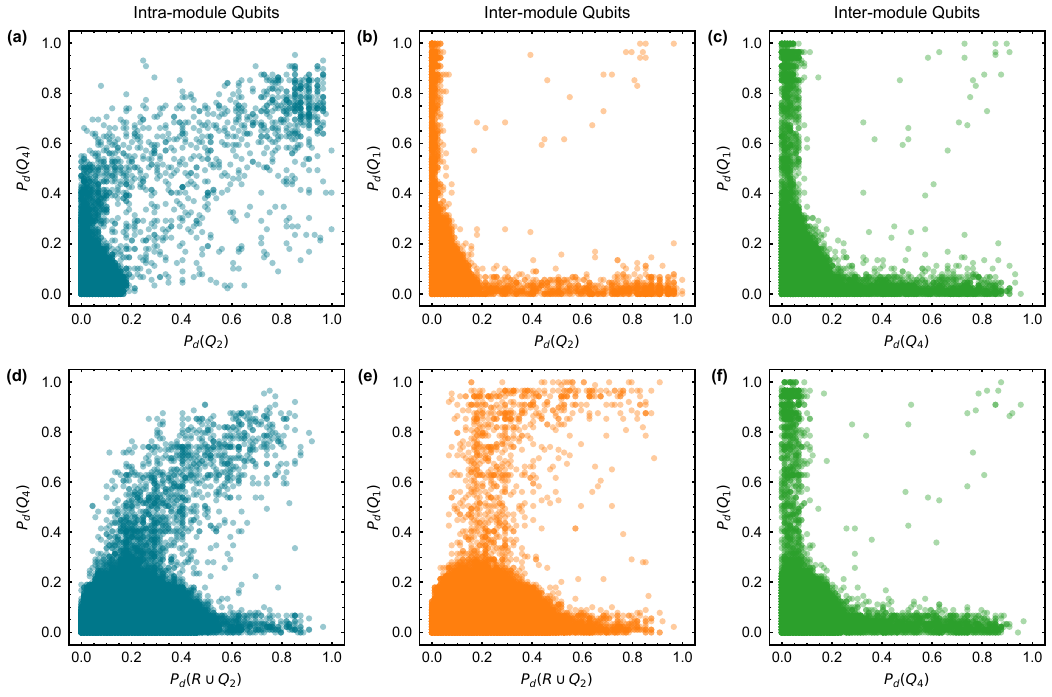}
    \caption{Scatter plots of two-qubit joint decay probabilities averaged over $N_b=100$ measurement cycles. (a), (b) and (c) Data for qubits $Q_2-Q_4$, $Q_1-Q_2$, and $Q_1-Q_4$ respectively, where $Q_2$ and $Q_4$ are located on the same daughterboard module while $Q_1$ is on the other daughterboard. (d) and (e) Scatter plots including decay probabilities for $Q_2$ serving as an ancilla for the motherboard router $R$, together with $Q_4$ and $Q_1$, respectively. (f) $Q_1-Q_4$ coincident data taken simultaneously with data in panels (d) and (e). In all panels, points close to the diagonal indicate correlated increases in decay probabilities in two qubits while points near the axes indicate decay probability increases for a single qubit or the router.}
    \label{fig:3}
\end{figure*}

\subsection{Time domain characterization of CRL events}
In Fig.~\ref{fig:2} we show examples of the time-dependent response of qubits $Q_2$ and $Q_4$. In panels (a) and (b) we show the decay probabilities taken in the same time span for qubits $Q_2$ and $Q_4$, showing a few coincident CRL decay events, and in panels (c) and (d) we show the detailed time dependence for one such CRL event, with the decay event onset in coincident time bins and somewhat different recovery time constants, $2.7$ ms for $Q_2$ and $6.1$ ms for $Q_4$. In Fig.~\ref{fig:2}(e), we display the cumulative distribution of fitted recovery time constants $\tau$ for the three qubits, with the shaded region representing $\pm \sigma$ (one standard deviation) in the fit recovery times. We find mean recovery times for $Q_1$ of $3.7\pm0.1$ ms, $Q_2$ of $2.2\pm0.1$ ms and $Q_4$ of $6.2\pm0.1$ ms; see Appendix \ref{sec:lifetime}. These results are consistent with measurements from other groups \cite{Harrington2024, Yelton2025}, which show longer recovery times when the qubit island-junction connection is via a thicker, lower superconducting gap metal lead, and the ground plane-junction connection is via a thinner, higher gap lead. In a separate set of measurements with longer $t_{\rm wait}$, we establish an average event interval of about $2$ minutes; see Appendix~\ref{sec:stat_event_arrival} for details.

\begin{figure*}[tb]
    \centering
    \includegraphics[width=\linewidth]{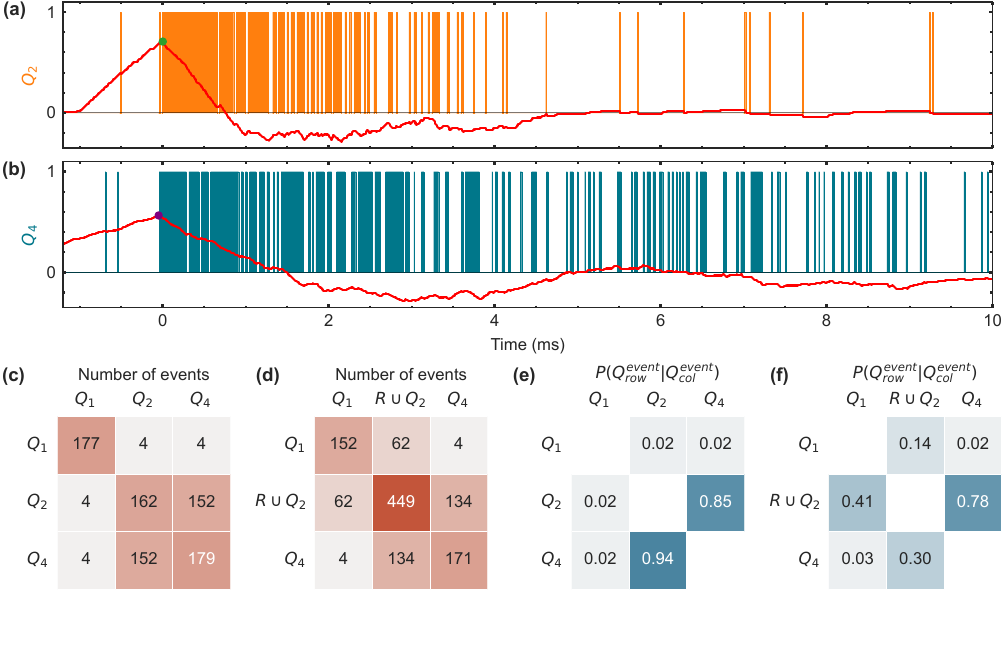}
    \caption{Event statistics. (a), (b): Short sequences of bit string data, post-processed to identify a decay error $ggg$ as $1$, otherwise as $0$, taken for a coincident CRL event in $Q_2$ and $Q_4$. The convolution of a template contrast function with these data is shown in red (see main text and Appendix~\ref{sec:start_times}), allowing identification of the event start time, indicated by a green (purple) point in (a) and (b). By using the threshold technique described in Appendix~\ref{sec:pp-filter}, we identify all peaks in the bit string data for each qubit and for the qubit-router measurements. We tabulate in (c) and (d) the total number of events in each qubit on the table diagonals and the number of coincident events between different qubits in the off-diagonal elements, where events are classified as coincident if they occur within $1$ ms of each other. (e), (f): Ratio of the number of coincident events in $Q_{\mathrm{row}}$ given an event in $Q_{\mathrm{col}}$, extracted from panels (c) and (d), respectively.}

    \label{fig:4}
\end{figure*}

\subsection{Correlated decay events}
To visualize correlations in the qubit decay probabilities, in Fig.~\ref{fig:3}(a)-(c) we display scatter plots of the joint decay probabilities for each pair of qubits. Points near the origin represent time bins in which both qubits had small decay probabilities, and mostly represent isolated SPAM errors. Points away from the origin but close to either axis indicate a CRL event isolated in one qubit. Points nearer the diagonal represent CRL events occurring in both qubits in the same time bins. Panels (b) and (c) show very few coincident events occurring in qubits on different modules, while panel (a) shows a significant number of such events occurring in both qubits on the same module, with very few CRL events in just one qubit. Events in $Q_2$ have a faster recovery time compared to $Q_4$, so $Q_4$ has more data points with larger probability values than $Q_2$, resulting in a bias towards the $Q_4$ axis. 

In Fig.~\ref{fig:3}(d) and (e), we display similar scatter plots for the decay probability $P_d$ of qubit $Q_2$ used as an ancilla for the router $R$, together with $P_d$ for $Q_4$ and $Q_1$, respectively. Note that when using $Q_2$ as an ancilla for $R$, elevated decay probabilities will occur when either $Q_2$ or $R$ experiences reduced $T_1$ times; events in daughterboard B thus cannot be unambiguously separated from those occurring only in the motherboard; we thus designate the probability as $P_d(R \cup Q_2)$. In panel (d), we see that decay events in $Q_4$ are highly correlated with those in $R \cup Q_2$, as expected, although the distribution is notably different from that in panel (a). Similarly in panel (e) we see larger decay probability events in $Q_1$ are highly correlated with those in $R \cup Q_2$. Finally in panel (f), we show the scatter for $Q_1$ and $Q_4$ with the same data as in panels (d) and (e), showing that events in the two daughterboards remain weakly correlated. 

\subsection{Event statistics \label{sec:correlation_statistics}}
We identify CRL events using a method described in Refs.~\cite{Harrington2024, McEwen2021, McEwen2024}. We post-process the raw data with a sliding template of $ggg$ as described in Sec.~\ref{sec:measurement_scheme}, recording a $1$ when observing this sequence and $0$ otherwise. Fig.~\ref{fig:4}(a), (b) show processed data for qubits $Q_2$ and $Q_4$, respectively, for a single coincident event. We cross-correlate these data with an exponentially decaying template function, and identify CRL events as those with a cross-correlation value above a certain threshold (see Appendix~\ref{sec:pp-filter}).

To identify event start times, we use a contrast template equal to $+1$ for $0~\mathrm{ms}< t < 1~\mathrm{ms}$, and $-1$ for $-1~\mathrm{ms}< t<0~\mathrm{ms}$, zero-valued elsewhere, providing a reasonable match with the representative data in Fig.~\ref{fig:4}(a) and (b). We find the convolution integral of the contrast template with the data at each time point, and identify the maximum of this convolution as the event start time, as represented by the green and purple dots in Fig.~\ref{fig:4}(a) and (b). We can compare event start times in different qubits; we identify coincident events as those where the start times differ by less than $1$ ms; more details can be found in Appendix~\ref{sec:iad}.

For the qubit-only measurements as in Fig.~\ref{fig:1}(c), we tabulate the total number of CRL events in each qubit $Q_1$, $Q_2$ and $Q_4$, displayed as diagonal elements in Fig~\ref{fig:4}(c); the number of coincident events in two qubits are shown as off-diagonal elements. For measurements involving the router monitored by $Q_2$ as in Fig.~\ref{fig:1}(d), the analogous data is tabulated in Fig.~\ref{fig:4}(d). We convert the off-diagonal elements to probabilities in Fig.~\ref{fig:4}(e) and (f), calculating the ratio of the number of off-diagonal events in $Q_{row}$ to the total events in $Q_{col}$. This number can be interpreted as the conditional probability ($P(Q_{row}|Q_{col})$) of observing a coincident event in $Q_{row}$, given an event occurs in $Q_{col}$.

From Fig.~\ref{fig:4}(e) we see that $85\%$ of the events in $Q_4$ are coincident with those in $Q_2$, but only $2\%$ are coincident with $Q_1$. Similarly, for all events in $Q_2$, $94\%$ are coincident with $Q_4$, but only $2\%$ are coincident with $Q_1$. These demonstrate a high degree of event correlation between qubits in the same module, but very little correlation between events on different modules. We can estimate the probability of time-correlated events occurring by chance in two qubits, as detailed in Appendix~\ref{sec:chance}, and find this probability is extremely small; the small number of observed coincident events in $Q_1-Q_2$ and $Q_1-Q_4$ are more likely due to a common cause, such as cosmic ray shower events.  For measurements involving the router, we observe a much higher number of total events, as router events are combined with events in $Q_2$ as mentioned earlier. We note also that the router circuit covers a larger area than any of the qubits, and the motherboard is also much larger than the daughterboards in area. In Fig.~\ref{fig:4}(f), $41\%$ of events in $Q_1$ are correlated with events in $R \cup Q_2$ (note from Fig.~\ref{fig:4}(e) the very small correlation between $Q_1$ and $Q_2$). This likely is due to high-energy particles passing through both daughterboard A and the motherboard, resulting in coincident events. This same mechanism can explain some of the coincident events between $Q_4$ and $R \cup Q_2$, although dominated by correlated events between $Q_4$ and $Q_2$.


\section{Discussion and Outlook}
We have measured CRL events in a modular superconducting quantum processor, using as event detectors three qubits in two flip-chip daughterboard modules and a router element located on an supporting motherboard. Identifying CRL events as prolonged periods of short energy relaxation times, each qubit exhibits on average a CRL event approximately every three minutes of observation time. We observe coincident CRL events that are strongly correlated between qubits on the same daughterboard module but only very weakly correlated for qubits on different modules. Events in the motherboard, as detected by the router, are well-correlated with events in one or the other module, consistent with events from high-energy particles traversing through one module and the motherboard in a mostly vertical direction. The localization of events to one module is straightforward to implement and compatible with many other CRL event mitigation techniques. It also indicates that error-correcting codes such as that proposed in Ref.~\cite{Xu2022} would allow preservation of the quantum state of a modular quantum processor suffering from such high-energy impacts, providing an architectural design, supplementing materials-based suppression of QP poisoning due to such events \cite{Pan2022, Loer2024, McEwen2024, Nho2025}.

\begin{acknowledgments}
Supported by the Army Research Office and Laboratory for Physical Sciences (ARO grant W911NF2310077), the Air Force Office of Scientific Research (AFOSR grant FA9550-20-1-0270 and MURI grant FA9550-23-1-0338), DARPA DSO (grant HR0011-24-9-0364), and in part by UChicago's MRSEC (NSF award DMR-2011854), by the NSF QLCI for HQAN (NSF award 2016136), by the Simons Foundation (award 5099) and a 2024 Department of Defense Vannevar Bush Faculty Fellowship (ONR N000142512032). Results are in part based on work supported by the U.S. Department of Energy Office of Science National Quantum Information Science Research Centers. We made use of the Pritzker Nanofabrication Facility, which receives support from SHyNE, a node of the National Science Foundation's National Nanotechnology Coordinated Infrastructure (NSF Grant No. NNCI ECCS-2025633). The authors declare no competing financial interests. Correspondence and requests for materials should be addressed to A. N. Cleland (anc@uchicago.edu).
\end{acknowledgments}

\appendix

\section{Device characteristics\label{sec:devparm}}
Here we provide parameters for the device used for the measurements reported here. These include full qubit characteristics and the lifetime of the router measured by swapping excitations from a qubit.
\subsection{Qubit parameters\label{sec:qubitparams}}
\begin{table*}[htbp]
 \caption{Qubit parameters. $\omega_{q}$ is the qubit idle frequency, $T_{1}$ ($T_{2}^{*}$) is the qubit energy (Ramsey) decay time, and $\mathcal{F}_{\mathrm{SQG}}$ is the average single-qubit gate (X and X/2) fidelity measured by randomized benchmarking. $\omega_{rr}$ is the readout frequency, and $F_{g}$ ($F_{e}$) represents the readout fidelity of $\ket{g}$ ($\ket{e}$), defined by initializing the qubit to the respective fiducial state then measuring the probability of the qubit in the prepared state. \label{tab:qubit_params}}
 \centering
 \begin{ruledtabular}
 \bgroup
 \def\arraystretch{1.5}
 \begin{tabular}{cccc}
   &$Q_{1}$&$Q_{2}$&$Q_{4}$\\
   \hline
   $\omega_{rr}/2\pi$&5.7630 GHz&5.6997 GHz&5.6753 GHz\\
   $\omega_{q}/2\pi$&4.5852 GHz&4.5409 GHz&4.6553 GHz\\
   $F_{g}$&99.7\%&99.6\%&98.8\%\\
   $F_{e}$&97.4\%&98.2\%&95.9\%\\
   $T_{1}$&36.0 $\mu \mathrm{s}$&41.4 $\mu \mathrm{s}$&11.6 $\mu \mathrm{s}$\\
   $T_{2}^{*}$&671 ns&965 ns&505 ns\\
    $\mathcal{F}_{\mathrm{SQG}}$&$(99.52\pm0.05)\%$&$(99.71\pm0.10)\%$&$(99.58\pm0.15)\%$
 \end{tabular}
 \egroup
 \end{ruledtabular}
\end{table*}
Three qubits $Q_1$, $Q_2$, and $Q_4$ are used in this experiment, and their parameters are summarized in Table~\ref{tab:qubit_params}. We note that the wiring used in this experiment is identical to that in Ref.~\cite{Wu2024}.

\subsection{Router lifetime\label{sec:router_lifetime}}
\begin{figure}[tb]
    \centering
    \includegraphics[width=\linewidth]{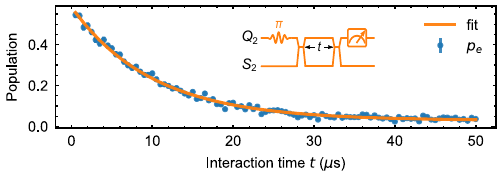}
    \caption{Router $T_1$ measurement. We measure the lifetime of the router by performing qubit-coupler swaps and varying the delay between two swap operations, as shown inset. A fit to an exponential decay yields the router $T_1=10.1~\mu\mathrm{s}$.}
    \label{fig:router_T1}
\end{figure}

To detect CRL events in the motherboard, we swap excitations from qubit $Q_2$ into the switch $S_2$ that forms an integral part of the router $R$. We use the same method to measure the router energy lifetime, shown in Fig.~\ref{fig:router_T1}, from which we arrive at a router lifetime of $T_1 = 10.1~\mu\mathrm{s}$. We note that in Fig.~\ref{fig:router_T1}, the population swapped from the router to the qubit is relatively small, mainly due to the large coupling strength, greater than $100~\mathrm{MHz}$, between the qubit and the router. Small distortions in the flux control can then lead to non-trivial leakage and imperfect swaps. We also note that as the switch $S_2$ is biased to its idle point after the initial swap with $Q_2$, the switch excitation is distributed among all four switches in the router before we perform the return swap to $Q_2$, so decoherence in any of the switch elements can affect the router lifetime. This also applies to the observation of CRL events in the motherboard, which means the events detected by $Q_2$ should represent a collective effect of the entire router footprint, rather than the switch $S_2$ itself.

\section{Time domain characteristics of CRL events}
Here we provide more details about CRL event lifetimes as measured for different qubits, as well as the statistics of the correlated event arrival times.

\begin{figure}[b]
    \centering
    \includegraphics[width=\linewidth]{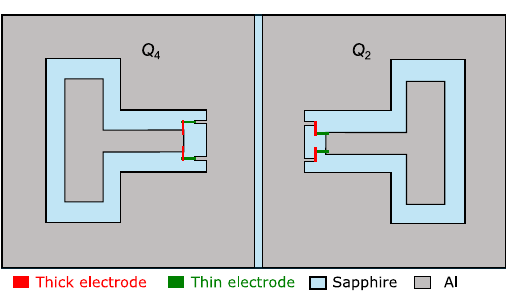}
    \caption{Schematic representing daughterboard B, showing junction orientations with respect to qubits. Not to scale, and only showing the relevant circuit elements.}
    \label{fig:jj}
\end{figure}

\subsection{Event recovery lifetimes\label{sec:lifetime}}
We deposit the qubit Josephson junctions via a Dolan bridge method~\cite{Dolan1977}. The first angled deposition of Al is $32.5~\mathrm{nm}$ thick, and the second deposition, also Al, is $100~\mathrm{nm}$ thick. This creates an asymmetry in the junction structure, where the thicker top Al electrode is expected to have a lower superconducting gap than the thinner base Al electrode~\cite{Marchegiani2022}. To tunnel through the junction from the top to the base electrode, QPs then need to obtain energy from the environment; this is the main source of premature qubit relaxation. As shown in Fig.~\ref{fig:jj}, due to differences in fabrication layout, for qubits $Q_1$ and $Q_2$, the qubit capacitor is connected to the thinner junction electrode, with larger gap, and the ground plane is connected to the thicker junction electrode, with smaller gap. However, in $Q_4$, this is reversed. We can then infer that decay events in $Q_1$ and $Q_2$ are caused by QPs tunneling from the ground plane side of the qubit, whereas decay in $Q_4$ is caused by QPs tunneling from the qubit capacitor side of the qubit. The longer recovery times shown in Fig.~\ref{fig:2}(e) for $Q_4$ compared to $Q_1$ and $Q_2$ are perhaps due to the greater diffusive volume of the ground plane compared to the much smaller qubit capacitor volume.

\begin{figure}[tb]
    \centering
    \includegraphics[width=\linewidth]{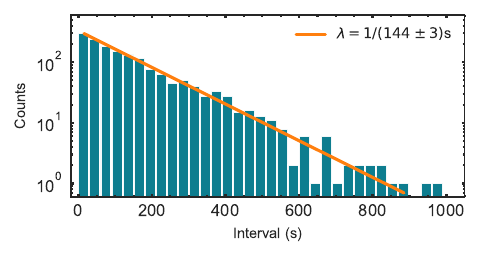}
    \caption{Histogram of CRL event intervals, compared to a fitted Poisson distribution (orange dashed line) with an average event occurrence rate of $1/(144\pm 3~\mathrm{s})$.}
    \label{fig:fig_event_interval}
\end{figure}

\subsection{Statistics of event arrival times\label{sec:stat_event_arrival}}
If the sources of CRL events are uncorrelated, the intervals between adjacent events will obey Poisson statistics. To probe this, we measure CRL event intervals using the same sequence as in Fig.~\ref{fig:1}(c), but with a much longer wait time $t_\mathrm{wait} = 200~\mu\mathrm{s}$. This increases the ratio of detector ``on'' versus ``off'' times, and as this is much longer than the qubit lifetimes, the qubits are passively reset to their ground states before the end of each cycle. We average the raw data into bins of five measurements, this time however without post-processing by template-matching, then search for event peaks in the same manner as in the main text. We then build a histogram of adjacent event intervals, shown in Fig.~\ref{fig:fig_event_interval}. The probability distribution of the event intervals appears Poissonian, from which we can extract an event occurrence rate of around $1/(144\pm 3~\mathrm{s})$. 

\subsection{Probability of coincident events by chance}
\label{sec:chance}
The probability of two uncorrelated CRL events occurring in two qubits within a $\Delta t = 1$ ms sampling window can be estimated from the Poisson distribution:
\begin{equation}
\begin{aligned}
    P(Q_j)P(Q_k) &= \left [ e^{-\lambda \Delta t}(\lambda \Delta t) \right ]^2
    \\ &\approx 4.82 \times10^{-11},
\end{aligned}
\end{equation}
with $\lambda= 1/{144~\mathrm{s}}$. With a total measurement time of $5.87$ hours (corresponding to $\approx 2 \times 10^7$ individual $1$-ms time samples), the expected total number of coincident events is $\sim 10^{-3}$. Compared to the off-diagonal elements in Fig.~\ref{fig:4}(c), it appears the coincident events are likely due to an underlying common cause, such as cosmic ray showers or other correlated CRL events.

\begin{figure}[tb]
    \centering
    \includegraphics[width=\linewidth]{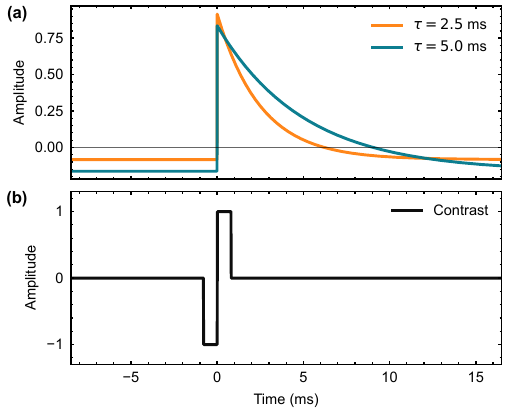}
    \caption{(a) Filter function for data cross-correlation to find all qualifying events. (b) Contrast template function, spanning $200$ measurement cycles between ($-1$ ms, $1$ ms), used to identify event arrival times.}
    \label{fig:filter}
\end{figure}

\section{Event identification analysis}
Here we discuss how we identify CRL events in each qubit and determine the event arrival times.
\subsection{Matched filter\label{sec:pp-filter}}
In order to identify CRL events in the qubits, we follow a scheme similar to that of Refs.~\cite{McEwen2021, Harrington2024, McEwen2024}. We use the template filter function shown in Fig.~\ref{fig:filter}(a), with a sudden onset, followed by an exponential decay, with baseline offset from zero in order that the integral over a preset time interval is zero, given by
\begin{equation}
    f(t) = 
    \begin{cases} 
     c, & t < 0, \\
     c + a e^{-t/\tau}, & t \geq 0.
    \end{cases}
    \label{eq:piecewise}
\end{equation}

\begin{figure}[tb]
    \centering
    \includegraphics[width=\linewidth]{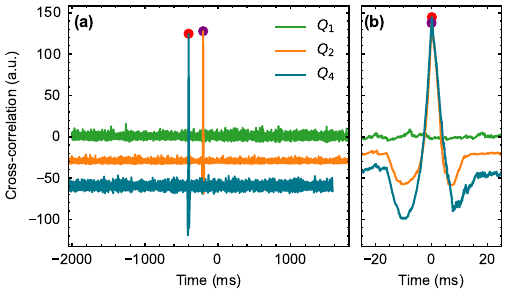}
    \caption{(a) Cross-correlation of the matched filter function with template-matched raw data for the three qubits, in the vicinity of a $Q_2-Q_4$ coincident event. Each trace is offset by $100$ ms along the time axis and 20 units along the vertical axis for better visualization. (b) Shorter time detail of the overlapping cross-correlated data.}
    \label{fig:cc}
\end{figure}

To process the raw data,  we post-process the qubit measurements by applying a template of three consecutive ``$g$''s ($ggg$), generating a result as shown in Fig.~\ref{fig:4}~(a) and (b). We find a sliding window cross-correlation of this result with the matched filter function, where we use a filter function with $\tau = 2.5~$ms for $Q_1$, $Q_2$ and $\tau = 5~$ms for $Q_4$, in accordance with the data in Fig.~\ref{fig:2}(e). We always choose the amplitude $a=1$, and the background $c$ is chosen to make the integral of the filter function, over the 24 ms interval shown, equal to zero.

We classify any instances where the value of this cross-correlated function goes above $25$ a.u. as an event candidate. The choice of this value is such that it is roughly $4\sigma$ away from the center of the peak clusters, where $\sigma$ is the maximum observed standard deviation of values visually confirmed to be outside the ``peak'' region across all data. Fig.~\ref{fig:cc} shows an example event, with a sharp contrast from the prevailing background.

\begin{figure}[tb]
    \centering
    \includegraphics[width=\linewidth]{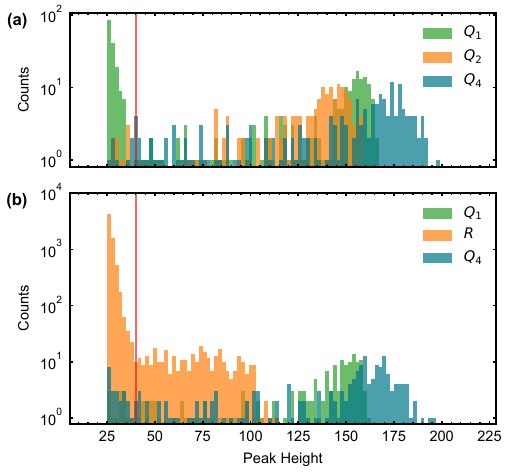}
    \caption{Peak heights calculated by cross-correlating the matched filter with template-matched data for (a) qubit-only measurements and (b) qubit-router measurements. Red lines are thresholds above which we consider a CRL event.}
    \label{fig:pk_hts}
\end{figure}

We investigate the statistical distribution of peak heights for all event candidates, shown in Fig.~\ref{fig:pk_hts}. For the qubit-only measurements in panel (a), we see a dropoff in peak heights for $Q_1$ immediately after the event candidate cutoff of 25, suggesting that these peaks are likely due to greater SPAM errors in $Q_1$.  The rest of the peak heights are clustered around $150$ for all three qubits, which are the peaks arising from CRL events. For the qubit-router measurements in panel (b), we see a similar initial drop followed by a more uniform distribution for the router $R$; and for $Q_1$ and $Q_4$, we see a cluster of peaks at values around $150$, as before. We choose a threshold peak height of $40$ a.u. as an event threshold, considering all peaks above this value as CRL events.

\subsection{Event start times\label{sec:start_times}}
We identify event start times by convolving the post-processed data with a contrast template $c(t)$, shown in Fig.~\ref{fig:filter}(b), defined as
\begin{equation}
    c(t) = 
    \begin{cases} 
     -1, \quad&-\Delta t\leq t < 0, \\
     +1, \quad&0\leq t \leq  \Delta t.
    \end{cases}
    \label{eq:contrast}
\end{equation}

Here $\Delta t=1~\mathrm{ms}$ is chosen to obtain a good signal-to-noise ratio. We define the event start time as the point at which the convolution integral reaches its maximum, as illustrated in Fig.~\ref{fig:4}(a) and (b). 

\subsection{Event inter-arrival delays}
\label{sec:iad}
\begin{figure}[tb]
    \centering
    \includegraphics[width=\linewidth]{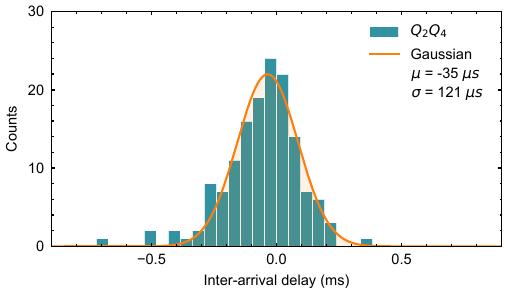}
    \caption{Distribution of event delays between $Q_2$ and $Q_4$, showing a high degree of coincidence.}
    \label{fig:iad}
\end{figure}

Using the start time information, we calculate the event inter-arrival delays between qubits, as mentioned in Sec.~\ref{sec:correlation_statistics}. In Fig.~\ref{fig:iad}, we display a histogram of event delays between $Q_2$ and $Q_4$, located on the same daughterboard, showing that most coincident events occur within $\pm 0.5$ ms, with a Gaussian fit yielding a mean of $-35~\mu\mathrm{s}$ with a standard deviation of $121~\mu\mathrm{s}$. As there are only a few coincident events between $Q_1- Q_2$ and $Q_1-Q_4$, these are not shown here. 

\clearpage
\bibliography{ref}
\end{document}